# Measurement of b quark EW couplings at ILC

S. Bilokin, R. Pöschl and F. Richard.

Laboratoire de l'Accélérateur Linéaire (LAL), Centre Scientifique d'Orsay, Université Paris-Sud XI, BP 34, Bâtiment 200, F-91898 Orsay CEDEX, France

______________________________________________________________________

**Abstract**: This paper describes an analysis performed at 250 GeV centre of mass energy for the reaction e+e- -> $b\bar{b}$ with the International Linear Collider, ILC, assuming an integrated luminosity of 500 fb$^{-1}$. This measurement requires determining the b quark charge, which can be optimally performed using the precise micro-vertex detector of the detector ILD and the charged kaon identification provided by the dE/dx information of its TPC. Given that the forward backward asymmetry is maximal for e$_{-L}$ (Left-handed electron polarisation), it has been necessary to develop a new method to correct for unavoidable angular migration due to b charge mis-measurements. This correction is based on the reconstructed events themselves without introducing external corrections which would induce large uncertainties. With polarized beams, one can separate the Z and γ vector and axial couplings to b quarks. The precision reached is at the few per mill level, and should allow to confirm/discard the deviation observed at LEP1 on the ZbRbR coupling. Model independent upper bounds on the tensor couplings, $F_{2V}$ and $F_{2A}$, are derived using the shape of the angular distribution.

## I.    Introduction

So far, LEP1 has determined the b quark couplings to the Z boson by measuring the b partial width and the forward-backward asymmetry called AFBb. These quantities provide the most precise value of s²w at LEP1. It turns out that this value is at three standard deviation away from the very precise value from SLD using beam polarisation.

Redoing precisely this measurement is therefore a priority for future e+e- colliders. This is possible at the Z pole, preferably with beam polarisation and very precise b tagging.

In this paper, we intend to prove that the International Linear Collider [1], with polarized beams and high luminosity, offers a unique opportunity for precise measurements well above the resonance, where both Z and photon exchanges are present. This additional complexity may turn up to be of great advantage since it allows, through γ-Z interference, to be sensitive to the sign of Z couplings and fully solve the LEP1 puzzle in an unambiguous way. Recall that the LEP1 anomaly can be interpreted up to a sign ambiguity for what concerns the right-handed coupling Zb$_R$b$_R$, referred hereafter as g$_{RZ}$, which shows the largest deviation [2].

At a centre of mass energy of 250 GeV, with $AFB_L$=0.70, the angular distribution $e^-_L e^+ \to b\bar{b}$ is maximally forward peaked, going almost to zero in the backward hemisphere, meaning that this measurement can be very sensitive to any BSM physics contribution populating this empty region. In contrast, for $e^-_R$, with $AFB_R$=0.28, this distribution is much flatter. The issue here is to prove that collecting a 500 fb$^{-1}$ luminosity one has sufficient accuracy to confirm or discard the LEP1 effect.

In contrast to the top quark case [3] where the quark decays before dressing up into a meson, the b quark mostly dresses up into scalar mesons before decaying and therefore cannot provide an adequate tool to separate left and right chiralities. Beam polarisation is therefore indispensable to separate the two components and provide the right handed coupling to the Z boson $g_{RZ}$ which shows the largest deviation from the SM.

To extract the angular distribution of the b quark – resulting in the standard AFBb observable – one needs to determine the b quark charge. At LEP1 [4] this was done using the leptonic decays with a very limited efficiency and an important dilution due, in particular, to large mixing effects from neutral B mesons. The use of B charge determination, based on secondary vertices measurements, was also implemented but with reduced purity and efficiency given the limited accuracy due to the large radius of the beam pipe. In contrast, SLD could perform a superior charge determination [4] but was statistically limited.

In the present work, we intend to provide a precise and clean angular distribution that carries much more information than just AFBb, in particular in the backward hemisphere. As will be discussed in section II.2, this analysis is very demanding in terms of purity and requires tagging of both B decays, keeping an efficiency above 10% by combining two signatures. The most obvious one uses charged B mesons selected by reconstructing all secondary tracks measured in the micro-vertex detector. In addition, the dE/dx information from the TPC of ILD [5] allows to select charged kaons, which results in a substantial increase of the efficiency without losing purity.

As already mentioned, the angular distribution of the b quark in the process $e^-_L e^+ \to b\bar{b}$ is very peaked towards $\cos\theta_b$=1 and goes almost to zero at $\cos\theta_b$=-1. This feature therefore requires a very pure charge tagging to limit migration effects due to wrong b charge assignments. Even with the double tagging method, this purity is insufficient and a powerful method has been implemented to correct for the remaining b quark wrong charge assignments. This method, which will be explained in section II.3, uses the amount of events with contradictory signatures, for instance $B^+B^+$ or $B^-B^-$, to deduce the migration probability. It can be applied differentially at the various $\cos\theta_b$ bins. It starts directly from the data themselves without using external inputs which are too uncertain to provide a reliable correction to real data.

The large integrated luminosity available at ILC gives rates comparable to those accumulated by LEP at the Z resonance and requires a careful estimate of instrumental errors. A detailed analysis goes beyond the scope of the present paper but a preliminary estimate will be provided using the methods developed at LEP1 [4].

In a first stage [6] ILC will be able to measure this asymmetry at 250 GeV, collecting 500 fb$^{-1}$. The present paper is focused on that phase. The same measurement can be done at 500 GeV with similar aspects not treated here. There will be a high luminosity phase [6], allowing for more precise measurements.

Many aspects of the present work, like the detector performances, the generators and the definition of form factors and couplings and the references are common to the top quark analysis [3] and we do not intend to repeat them here.

Results of our analysis will be presented in section III and IV, while section V will briefly discuss possible interpretations. An Appendix is devoted to present the formulae used in this analysis.

# II. Analysis

The detector properties and the various tools used in ILD analyses are described in [3]. One uses a full simulation of the ILD detector, taking into account superimposed $\gamma\gamma$ events and apply full reconstruction. The results here are based on ilcsoft-v01-16-p10. The generator WHIZARD version 1.95 is used for the generation of signal and background samples. Beam spread induced by the bremsstrahlung process is included. $\gamma\gamma \rightarrow$ hadrons background events are also taken into account [3].

Initial state radiation is implemented as well as gluon radiation but NLO electroweak corrections are ignored. In the future, they will be needed given the high accuracies achieved.
Below are listed the relevant signal and background cross sections, that is for purely hadronic processes containing at least a pair of b quarks. The subscripts 'unpol', L and R indicate the polarisation configurations: unpolarised, left-handed electron beam and right-handed positron beam (L) and vice versa (R), with both beams fully polarized.

*Table 1: Basic cross sections at 250 GeV for processes producing at least a pair of b quarks. L(R) refer to the electron polarisation, assuming that e+ has a R(L) polarisation.*

| Channel | $\sigma$unpol  fb | $\sigma$L  fb | $\sigma$R  fb |
|---|---|---|---|
| $b\bar{b}$ | 1756 | 5629 | 1394 |
| $\gamma b\bar{b}$  (Z return) | 7860 | 18928 | 12512 |
| ZZ  hadronic with $b\bar{b}$ | 196 | 549 | 236 |
| HZ  hadronic with $b\bar{b}$ | 98 | 241 | 152 |

## II.1. Event pre-selection

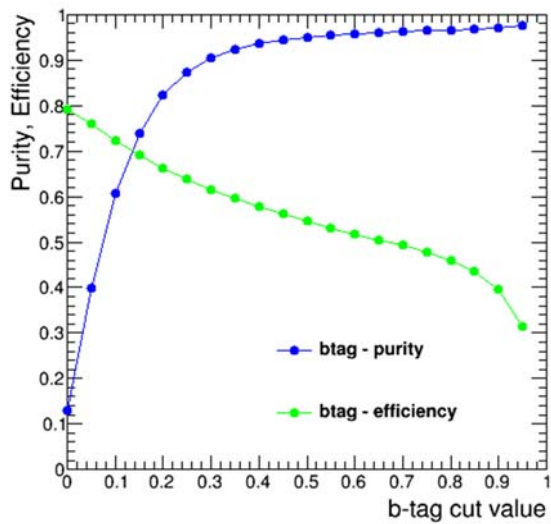

*Figure 1: Purity (blue curve), efficiency (green curve) versus the b-tag cut value for e-$_L$ polarisation.*

As for the top quark analysis, the b selection efficiency is related to tagging purity. Our analysis requires double b-tagging and therefore the contamination from charm production is negligible. The remaining contamination comes mainly from ZH and ZZ. This background is reduced using a selection against massive jets without a significant reduction in efficiency. One select events for which the sum of the masses of the two jets is below 150 GeV and with a sphericity S below 0.3. To avoid any angular dependence of the mass cut, one needs to correct for a moderate angular dependence of the energy response of the detector.
The mass selection against the ZZ and ZH contaminations has another virtue: it eliminates genuine events with gluon emission at large angle. This feature contributes to improve the accuracy on the event angle reconstruction.
Another issue, absent for top quark analysis, is the large probability for a radiative return to the Z pole. In most cases the ISR (Initial State Radiation) photon remains in the beam pipe but is easily eliminated by selecting visible masses above 180 GeV. In some cases the ISR photon appears inside the detector and can be identified when sufficiently energetic, above 40 GeV. Both cuts consistently select $b\bar{b}$ masses between ~200 GeV and 250 GeV. This selection corresponds to a 10% correction with respect to the Born cross section. For AFB, this correction is very much reduced: negligible for e-$_L$, it is ~1% for e-$_R$.

Figure 1, valid for e-$_L$, shows the purity versus efficiency versus the b-tag cut value.

## II.2 b quark charge selection methods

### II.2.1 B charge selection based on micro-vertex tracking

The most straightforward method to identify the b quark charge is to select charged B mesons which are unaffected by mixing and therefore provide a pure signature. Due to tracking inefficiency, this method suffers from contaminations from neutral mesons where one track is lost. To estimate this effect one should recall that default track inefficiency is at the level of 10% [7] and therefore had to be improved for this analysis.

The TPC detector reconstructs 99% of the tracks meaning that the 10% inefficiency corresponds to tracks which have been found but not selected for reconstructing secondary vertices. Part of them seem non recoverable since they do not have a significant offset with respect to the main vertex. Recovery procedures have been implemented, resulting in a reduction from the initial 10% to the present 5.8%. Improvements are under implementation and this analysis serves as a benchmark for these developments [7]. In the barrel region about 45% of these lost track do not seem recoverable since they have no significant offset with respect to the primary vertex, which is a prerequisite for the standard algorithms. The rest, 55%, are suffering from bad associations between detectors and can, in principle, be recovered.

For $|\cos\theta_b|>0.9$, the barrel micro-vertex detector becomes inefficient in the forward region covered by Si discs. These start at ~20 cm from the interaction point and therefore, due to multiple scattering in the material of the detector, the provided accuracy is insufficient for average momenta. Modifying this geometry and/or prolonging the 1$^{st}$ layer of the barrel is under consideration.

After applying the recovery procedure described in [7] one obtains a good correlation between the expected number of charged tracks connected to the secondary vertex and the reconstructed number: 66.5% of our events sit on the diagonal (see figure). This result is easy to understand given the average multiplicity of secondary tracks, ~6, which are involved in secondary vertices. The contamination for charged B mesons comes from neutral B mesons where one track is lost however this happens randomly and therefore only in half of the cases one gets a wrong signature. This feature explains why the average purity is p=0.84. This purity is insufficient for our purpose, hence the need for double tagging.

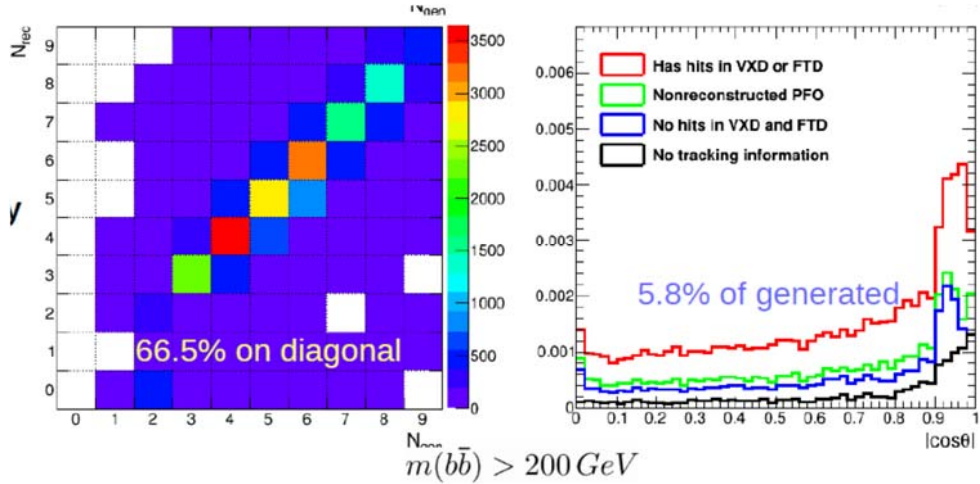

*Figure 2: Left figure shows the correlation between the number of secondary tracks generated and the number of tracks reconstructed. Right figure shows the angular distribution of the various categories of tracks missing for vertex reconstruction.*

Given that charged mesons occur in 40% of the cases, the double tagging requirement gives an efficiency below 16%. Including inefficiency limitations due to low B lifetime, this efficiency drops below 10%.

## II.2.2 B charge selection based on charged kaons

Charged kaons turn out to give a similar purity, at the ~80% level, as can be seen from the PDG and as confirmed by our generator. This figure may sound surprisingly good but can be intuitively understood recalling that a b quark decays into $cW^-$. The quark c gives an s quark, therefore a $\bar{u}s=K^-$ while $W^-$ can go into $\bar{u}s$, hence preferentially also into $K^-$.

Having ~0.8 charged kaons and ~3.6 charged pions per B decay, the selection requested from the TPC dE/dx appears easy.

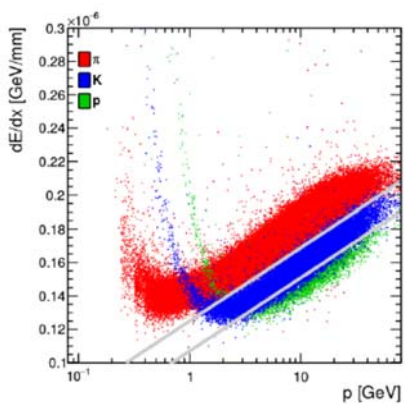

*Figure 3: Momentum dependence of dE/dx TPC deposit for pions, kaons and protons after applying an angular correction. The two lines define our selection.*

The TPC is used to identify charged kaons with a good efficiency down to $|\cos\theta_b|=0.9$ for momenta above ~3 GeV. Figure 3 is obtained after correcting for an angular dependence of dE/dx [8] and shows a clear separation between kaons, protons and pions for p>3 GeV. Kaons are identified with a purity reaching 97% with an 87% efficiency.

The TPC information does not allow for a full separation between charged kaons and protons. These particles have a 0.13 multiplicity per B decay. Those coming from B baryons, when confused with charged kaons, provide the wrong b charge assignment. For these reasons the resulting purity is p=0.79. We also allow for more than one charged kaon taking the overall charge: $K^-K^-$, $K^-K^-K^+$ are accepted while $K^+K^-$ is rejected.

We accept the combination with each jet containing kaons with opposite charges, and B⁺B⁻ events. This strategy results in a global efficiency of 13% for both polarisations with a purity of 97% for $e_{-L}$ and 95% for $e_{-R}$.

## II.3. Correcting for migration effects.

An angular migration occurs when both b quark charges are wrongly measured. This mistake amounts to flipping the value of $\cos\theta_b$ into $-\cos\theta_b$ and turns out to be unacceptable when this distribution is strongly peaked as is the case for left-handed electron polarisation $e_{-L}$. This is shown in figure 4 where the green distribution is the ideal one, while the black points are the result for the genuine reconstruction. To cope with this effect, we have used the large sample of events with contradictory charges. For example events with ++ and - - charges. Assume that around a certain value $|\cos\theta_b|$ we have rejected Nr events as being contradictory, Na+ accepted events as having $\cos\theta_b>0$ and Na- events with $\cos\theta_b<0$.

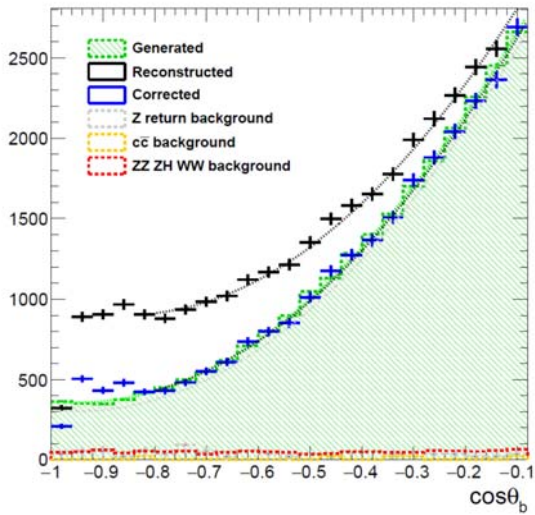

Figure 4: Backward hemisphere angular distribution for e-Le->bb showing the effect of our correction on reconstructed data. The red distribution corresponds to the remaining background.

One can write that:

Nr=2pqN where N is the total number of selected events in the $|\cos\theta_b|$ bin N=Na+ + Na- + Nr and p is the probability for a correct assignment at a given angle and q=1-p, the probability for a wrong assignment.

One then has the following two equations:

Na+=p²N+ + q²N-
Na-=p²N- + q²N+

where N+ and N- are the true number of events with positive and negative $\cos\theta_b$.

Solving these simple equations, one reports the corrected numbers in figure 4 as the blue points which are in perfect agreement with the ideal distribution in green. Statistical errors are trivially computed.

In practice, the recovery procedure shows that the resulting $p(\theta_b)$ stays constant with angle except for the very forward region $|\cos\theta_b|>0.9$ where there is also a significant drop of efficiency. This region will be excluded in the final fit described below.

## III.  Results

Figure 5 shows the results obtained with the two beams fully polarized and an integrated luminosity of 250 fb⁻¹. The naïve **counting approach** to measure the cross section and the asymmetry AFBb suffers from the limited acceptance for $|\cos\theta_b|>0.9$. This could, in principle, be done by correcting for inefficiencies but it would introduce systematic uncertainties difficult to evaluate. Instead one can adjust these data for by the theoretical distribution $S(1+\cos^2\theta_b)+A\cos\theta_b$, selecting the safe angular domain $|\cos\theta_b|<0.8$. This shape neglects the SM contribution $T\sin^2\theta_b$ which, according to the formulae given in Appendix, is very small given the large γ factor.

While S is related to the total cross section, A is simply related to AFBb without requiring a significant correction. For e-$_L$, the fit gives AFBgen/AFBrec=100.7±0.62%.

This result is achieved by including in the generated distribution (in green) the ISR effect and by subtracting the remnant ZZ and ZH backgrounds in the reconstructed sample. Therefore all reconstruction effects, like migration and energy resolution, are included, which constitutes a remarkable achievement. For e-$_R$, one has AFBgen/AFBrec=104.9±2.25%.

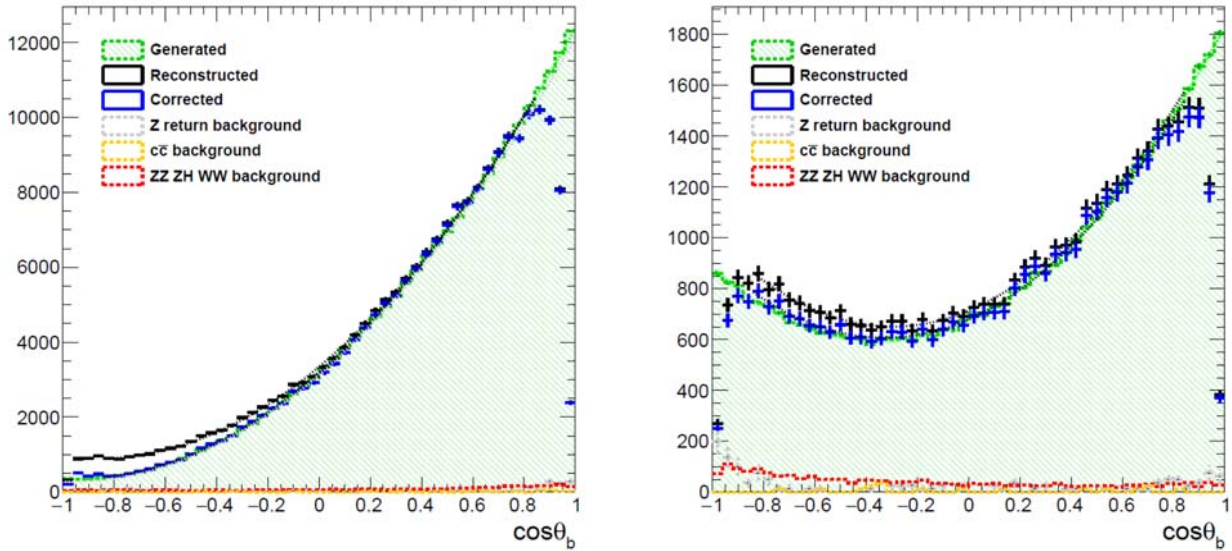

*Figure 5: Angular distributions obtained for e-L (left) and e-$_R$ (right) showing a perfect agreement of the corrected reconstructed data (blue) and the generated distribution (green). One assumes full beam polarisation for an integrated luminosity of 250 fb-1 for each case.*

In practice one will observe distributions with limited polarisation, P=∓80% for electrons and P'=±30% for positrons, therefore with a mixture of e-$_L$ and e-$_R$ chiralities. One will observe:

$$d\sigma_{-,+}/dcos\theta = 0.25[(R+L)(1-PP')+(P-P')(R-L)]=0.58L+0.035R \text{ and } d\sigma_{+,-}/dcos\theta = 0.58R+0.035L$$

where L=d$\sigma_L$/dcos$\theta$ and R=d$\sigma_R$/dcos$\theta$ (where L(R) refer to the polarisation of e-)

It is straightforward to take two combinations from the experimental plots which gives back pure L and R samples and which correspond to the curves above. One should take into account that the two samples will be taken with different luminosities. The sample e-$_L$ comes out with very small corrections given that the L cross section is four times the R cross section.

When doing this in detail, one finds that the naïve method which takes directly the two curves from figure 5 with appropriate polarisation and luminosity assumptions provides an excellent first order approximation. The largest correction to this approximation is observed for an e-$_R$ polarisation, with an increase of 11% on the errors of S and A.

A similar procedure can be followed for what concerns the ZZ and ZH backgrounds.

Our final selection provides 97% purity for e-$_L$ and 95% for e-$_R$. These figures clearly indicate that the error due to the background uncertainty should be very small. Since the ZZ and ZH cross sections will be precisely measured at 250 GeV, one can conservatively assume an uncertainty of 10% on the background contamination and therefore a 0.3% uncertainty for e-$_L$ (0.5% for e-$_R$) on the differential cross section measurement. This small number is larger than the errors coming from the luminosity and polarisation measurements of order 0.1% [3].

For the error on the b tagging efficiency similar considerations cannot be applied and therefore one needs to appeal to methods similar to those used at LEP1 which lead to errors at a few 0.1% level. Let us briefly recall what they were.

The basic idea is to extract the efficiency from the data themselves henceforth reducing the uncertainties due to the generator. For a given btag selection with an efficiency $\varepsilon_b$, one simply compares the amount of events with double btag, which is proportional to $\varepsilon_b^2$ to the amount of events with a single btag which is proportional to $\varepsilon_b$. From the ratio one extracts $\varepsilon_b$.

How do we compare to LEP1? For the left-handed polarisation e-$_L$, the amount of $b\bar{b}$ data is ~$10^6$, therefore comparable to one LEP1 experiment but, for a similar purity, our b jet tag efficiency is 40% instead of typically 20%.

Another significant advantage of ILC is the very small size of the beam spot position which allows to constrain the main vertex reconstruction. This uncertainty on the main vertex reconstruction turned out to be a severe source of correlation at LEP1. In simple terms, if the main vertex is wrongly reconstructed, the b tagging of the two jets are both affected and one cannot assume that their $\varepsilon_b$ are independent which is the prerequisite of the method. Recall that, at LEP1, the beam size was $\sigma y$~5μm and $\sigma x$~15μm while at ILC $\sigma y$~6 nm and $\sigma x$~0.6μm.

While it is beyond the scope of the present study to provide a precise number, from these arguments one can safely estimate that the uncertainty on $\varepsilon_b$ will be at least as good as for LEP1, that is ~0.2%.

For e-$_R$ the error on $\varepsilon_b$ will increase, following the rate, but it is legitimate to use the figure deduced from the e-$_L$ polarisation.

The running strategy at 250 GeV will be focussed on the HZ reaction and therefore the main part will be on e-$_L$ where this cross section is maximal for HZ. Therefore 67.5% of the luminosity will be spent in the combination e-$_L$e+$_R$, 22.5% on e-$_R$e+$_L$ and the rest on identical chiralities, which is only of use to measure the polarisation. Recall that e- is 80% polarized while e+ 30%.

In the first phase, 500 fb$^{-1}$ will be collected at 250 GeV with, later on, a high luminosity phase to reach 2000 fb$^{-1}$. Table 2 summarizes our numbers.

*Table 2: Relevant figures of our analysis.*

| Polarisations P P' | -80% +30% | +80% -30% |
|---|---|---|
| Efficiency % | 13 | 13 |
| Luminosity fb$^{-1}$ | 340 | 110 |
| Cross section fb | 3342 | 1012 |
| Background % | 3 | 5 |
| Syst % L+Pol+back+eff | 0.1+0.1+0.3+0.2 | 0.1+0.1+0.5+0.2 |
| Stat+syst % | 0.31+0.38 | 1+0.56 |

# IV. Determination of couplings

## IV.1 Two-parameter fit

The vector and axial couplings are usually deduced from AFB and the cross section, which are statistically uncorrelated. This counting method cannot be applied here, given the limited acceptance in the forward region. One can, instead, adjust the angular dependence by the expression:

$$S(1+ \cos^2\theta_b)+A\cos\theta_b$$

where S is related to the total cross section while A is related to the forward backward asymmetry AFB. The errors from the fit allow to deduce the errors on the various couplings. Our convention for the vector and axial couplings follow [9]. It is important to note that when AFB comes close to its maximal value, as is the case for the $e_{-L}$ polarisation, S and A parameters are almost fully correlated, which drastically reduces errors on axial and vector couplings which are obtained by combining both S and A parameters.

For $e_{-L}$, the statistical error on S and A, in relative values, are 0.31% and 0.38%. The correlation coefficient is $\rho$=0.84. For $e_{-R}$, these numbers become 1.0% and 3.9%, with a correlation of 0.30. This large increase simply reflects the fact that for $e_{-R}$ the cross section is ~4 times smaller and the collected luminosity is ~2 times smaller.

Since the correlation plays a major role in reducing errors on form factors, we have tried and succeeded to reproduce the result of the fit with a simple counting approach. One predicts a correlation coefficient, $\rho$=AFB, which comes pretty close to the results of our fits. One also predicts a ratio between the errors R=(dS/S)/(dA/A)=AFB, again in agreement with our findings. To fully take into account the fitting approach, one can use a likelihood method which says that $\rho$ is simply given by R, the ratio of errors, which comes closer to the result from the fit.

Knowing the errors on S and A and the correlation coefficients allows to compute all quantities for a given beam polarisation. If one needs to correlate quantities from the two polarisations one should be cautious in taking account the fact that instrumental effects are common to the two polarisations and therefore fully correlated.

Given the very small statistical errors obtained with $e_{-L}$, systematic effects should be carefully taken into account. In the previous section, the estimated relative error on the cross section, 0.38%, directly applies to S and A and with almost full correlation.

*Table 3: Summary of our results on the measurements of the relevant quantities defined in the text.*

| Quantity | $F_{1V\text{em}}$ | $F_{1VZ}$ | $F_{1AZ}$ | $\Re F_{2VL}$ | $\Re F_{2VR}$ | $g_{LZ}$ | $g_{RZ}$ |
|---|---|---|---|---|---|---|---|
| SM value | -1/3 | -0.41 | 0.593 | $2.3 \cdot 10^{-4}$ | $2.8 \cdot 10^{-5}$ | -0.42 | 0.077 |
| $\sigma$ Stat | 0.0015 | 0.0026 | 0.0031 | 0.0017 | 0.003 | 0.00135 | 0.00135 |
| Syst | 0.0005 | ~0 | 0.0013 | ~0 | ~0 | 0.00080 | 0.00027 |
| Sys+stat % | 0.45 | 0.63 | 0.56 | | | 0.37 | 1.76 |
| LEP1 % | | 1.0 | 2.4 | | | 0.36 | 8.2 |

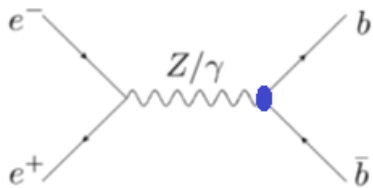

Given the the reaction $e_{-L}\ e+ \rightarrow b\bar{b}$, one can describe the b quark vertex part by the general expression [9]:

$$\Gamma_\mu^{Z/\gamma} = e[\gamma_\mu(F_{1V}^{Z/\gamma}+\gamma_5 F_{1A}^{Z/\gamma})+\frac{i\sigma_{\mu\nu}q^\nu}{2m_b}(F_{2V}^{Z/\gamma}+\gamma_5 F_{2A}^{Z/\gamma})]$$

Where $F_{1V\gamma}=Q_b$, $Q_b=-1/3$, $F_{1VZ}=(-1/4-Q_b s^2 w)/swcw$ and $F_{1AZ}=(1/4)/swcw$ are the usual form factors. One will also use the related coupling: $g_{LZ}=-1/2-Q_b s^2 w$, $g_{RZ}=-Q_b s^2 w$.

In the Appendix, one can see how $F_1$ and $F_2$ enter in the expressions of S and A, from which one can derive the errors given in table 3. LEP1 results were taken from [10].

An important conclusion can be drawn from the accuracy on $g_{RZ}$, recalling that the LEP1 anomaly predicts a deviation of ~25±10% on this parameter. Given the accuracy predicted at ILC, ~2%, this deviation should be either fully confirmed or definitely discarded. This is true even assuming that LEP1 was 'lucky' and the real effect is at the 10% level.

In this respect, it is worth noting that the luminosity sharing between $e_{-L}$ and $e_{-R}$ does not affect this result since about half of the error comes from $e_{-L}$. This sharing however affects the accuracy on $g_{LZ}$ which is slightly below the LEP1 accuracy.

For certain BSM schemes where a Z' only couples to $e_{-R}$ like in [11], the large error on S and A for $e_{-R}$ will translate into a poorer reach for these models.

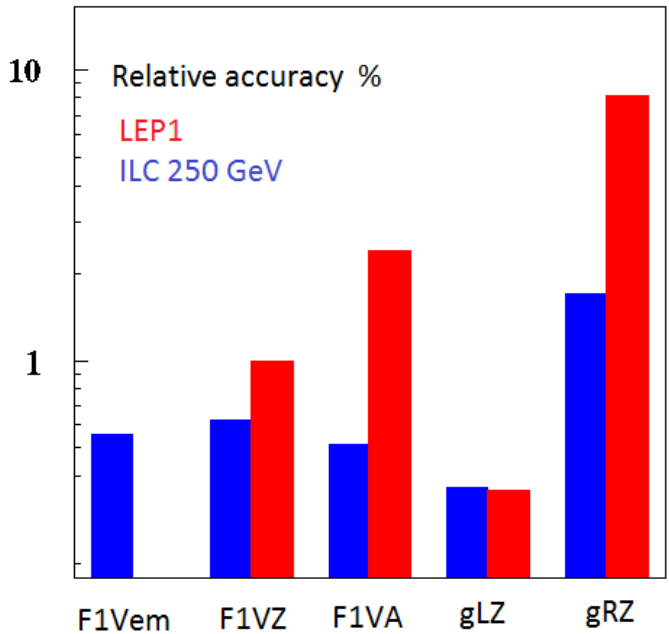

Figure 6: Comparison of relative accuracies achieved at LEP1 (in red) and those predicted for ILC (in blue) for a luminosity of 500 fb-1.

A few remarks are in order. For $e_{-L}$, the statistical accuracy on the cross section S and on the forward backward asymmetry are excellent, at the 0.3% level, and therefore come very close to the various uncertainties previously described: background subtraction, tagging efficiency and luminosity uncertainty. Adding quadratically these effects gives 0.38%. It is important to realize that these errors are fully correlated for what concerns the parameter S and A. Additionally, A suffers from residual migration effects which, given the perfect agreement observed after the corrections, appears negligible. As previously mentioned, the angular distribution could be affected by our mass cut against the ZZ and ZH backgrounds where there is an angular dependence of the energy resolution. This effect was indeed observed and corrected.

The present analysis is using a Whizzard version which comprises 1st order QCD corrections but has no EW corrections. These corrections are currently evaluated for the top quark sector and, contrary to expectations, appear to be quite large, especially for the $e_{-L}$ case where box diagrams with W exchange affect both the cross section and AFB [12]. It is hoped that such calculations will soon be implemented for the b quark.

Finally, table 3 also shows that accuracies are dominated by statistics and that with a high luminosity phase at 2000 fb$^{-1}$ one can improve on these errors by almost a factor 2. Systematics, in particular the tagging efficiency, will decrease with more data.

## IV.2 Three-parameter fit

For what concerns $F_{2V}$, the quoted accuracy in table 3 assumes, as usual, that only $F_{2V}$ is varied, while $F_{1V}$ and $F_{1A}$ remain constant. One is essentially sensitive, by interference effect, to the real part of $F_{2V}$. The SM values are very small and, for the QCD and QED parts, can be simply expressed in terms of a form factor $F_{2Q}$ such that $F_{2VL}=F_{1VL}F_{2Q}$ and $F_{2VR}=F_{1VR}F_{2Q}$ (see Appendix). For the EW and BSM contribution, this however is not true, since the couplings of exchanged particles depend on the chirality and therefore the form factor $F_{2Q}$ is different for bL and bR. For this reason we have chosen to give the experimental errors in term of $F_{2VL}$ and $F_{2VR}$.

To allow for a model independent extraction of $F_{2VL}$ and $F_{2VR}$, one should use the most general angular dependence of the cross section:

$$S(1+\cos^2\theta_b)+A\cos\theta_b+T\sin^2\theta_b$$

where, as can be seen in the Appendix, $T\sim|\gamma^{-1}F_{1V}+\gamma F_{2V}|^2+|\gamma F_{2A}'|^2$, showing that tensor couplings are enhanced given that for b quarks $\gamma\sim 25$. By performing a 3-parameter fit, one can therefore be sensitive to $F_{2V}$ and $F_{2A}$ without assuming that the vector terms do not vary, which means that one can have a full disentangling.

While this measurement is not sensitive to the SM QCD contributions which are very small, it can allow to measure BSM contributions which could, in principle, be much larger.

The 3-parameter fit gives an error twice as large on S, due to a strong correlation between S and T parameters. The error on A remains the same. As expected, the average value of T given by the fit is much smaller than for S, but the error on T is similar to the error on S. From this method, one finds, at the one standard deviation level, the following two bounds:

$$|F_{2VL}|<0.005 \quad \text{and} \quad |F_{2VR}|<0.011$$

Again, these bounds are model independent and do not make any assumptions about the behaviour of the vector terms. They are comparable to the size of errors obtained for $\Re F_{2VL}$ and $\Re F_{2VR}$ with the usual method (table 3). The 3-parameter fit also provides similar upper bounds on $F_{2AL}$ and $F_{2AR}$. It is however fair to say that a deviation on the T parameter cannot be unambiguously interpreted as the presence of CPV violation since it cannot be separated from an $F_{2V}$ contribution.

## V. Interpretations

### V.1 RS approach

In [2], two solutions are provided to interpret the effect observed at LEP1 on AFBb. From table 3, one concludes that both solutions can conclusively be tested. Varying the bR coupling by +25% is called RSb, by -225%, is called RSa. By measuring the angular distribution of e+e-> $b\bar{b}$ at 250 GeV one can visually distinguish between these two solutions.

With RSa there are deviations for both beam polarisations but it is worth noting that for left-handed electrons the measurement relies mainly on the backward hemisphere and therefore requires a clean elimination of the migration effect.

With scenario RSb, the difference with the SM is minute and confirming the effect seen at LEP1 is more challenging but numerically achievable with ILD accuracies. Visually, the difference is only significant for right-handed electrons.

How heavy are the new vector bosons responsible for these deviations? In the RS approach of [2], one assumes that there are extra symmetries, the so-called custodial symmetries, which allow to reduce the contributions to S and T parameters, meaning that KK resonances as light at 3-5 TeV are allowed. There are other approaches [13] which permit even lower masses.

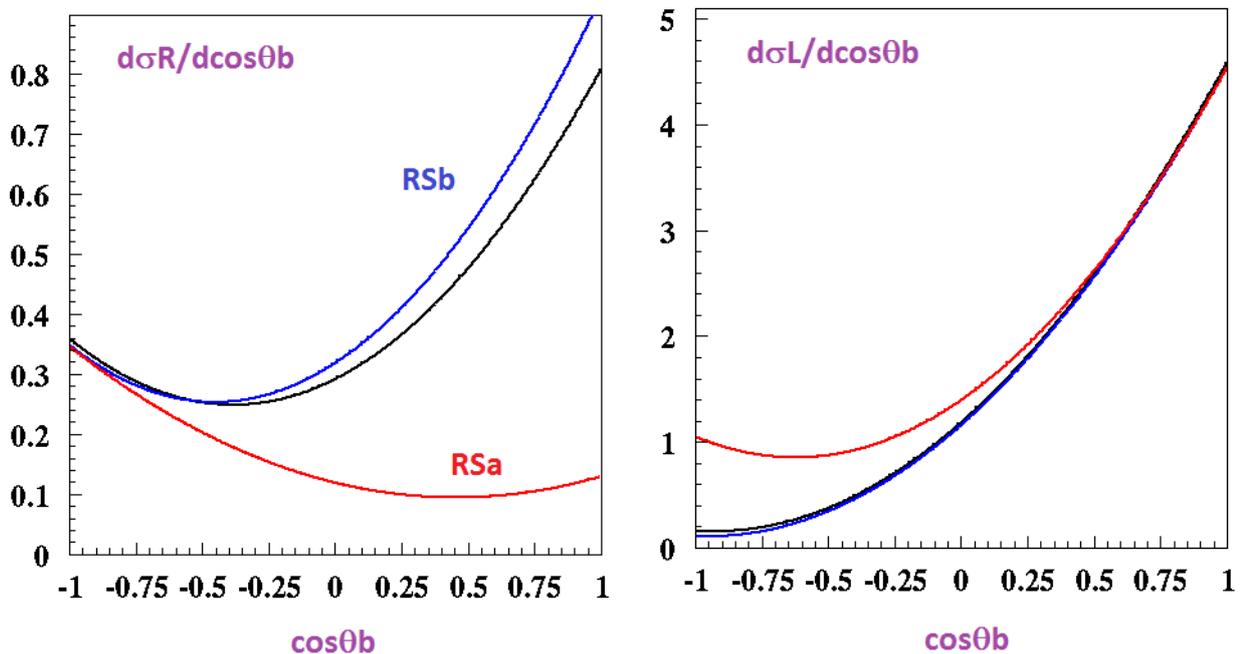

*Figure 7: Predicted angular distributions for ee->bb for the SM (black), the RSb solution (blue) and the RSa solution deduced from the LEP1 anomaly. Left is for $e_{-R}$ and right is for $e_{-L}$.*

So far, RS models imply that only the heavy quarks, top and bottom, are affected by the presence of the KK particles. In [11], which uses Higgs-vector boson unification scheme, all flavours, including leptons, are affected. Our method, based on charged kaon identification, would then be relevant to the charm quark leading, at first sight, to similar accuracies.

## V.2 Effective field theory

Instead of using an explicit interpretation, one can use an effective field theory approach, so-called EFT, which turns out to be simple to implement as discussed in [14]. This reference underlines that b quark EFT terms are closely related to t quarks terms by the simple fact that these quarks belong to the same EW doublet.

This is well illustrated by figure 8 where LEP1 measurements on Z$b\bar{b}$ are used to narrow down the ILC measurements on top quarks.

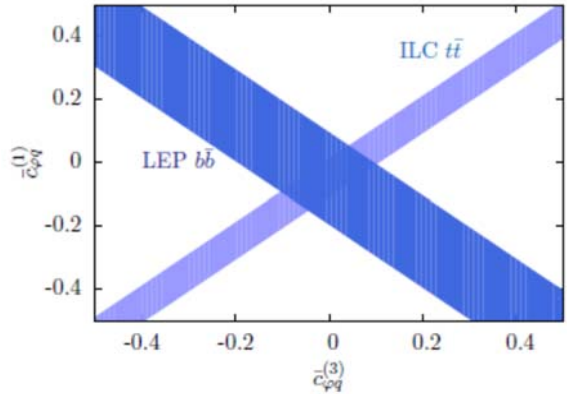

Given that ILC will provide more precise data than at LEP1, these figures will still be improved and it will become possible to extend the reach on new mass scales beyond 10 TeV.

*Figure 8: Expected accuracies on EFT coefficients using ILC top measurements (light blue) and including LEP1 measurements (dark blue).*

## VI. Summary and conclusion

Using a realistic set up and a full detector simulation and reconstruction, a comprehensive study of the reaction e-e+-> $b\bar{b}$ at √s=250 GeV was presented. Many aspects required extending our previous work for the top quark, in particular the need to improve b quark charge measurements and its challenges were thoroughly discussed.

With left-handed polarisation of the electron beam, the angular distribution is forward peaked at almost the maximum value and, even with the excellent vertexing properties of the ILD concept, unavoidable quark migration effects cause severe alterations of the backward angular distributions. A new method has therefore to be developed which relies on the use of the substantial population of events with like-sign charges. This correction methods is particularly robust since it makes no use of external information.

To improve the efficiency, charged kaons were used and resulted in a sample of events with similar purity. This establishes the interest of using charged kaon identification to improve electroweak measurements. This method can be applied to charm quarks which opens good prospects to test other models [11].

During this analysis, various weak points of the ILD detector were identified, in particular the drop in efficiency of b tagging for $|\cos\theta|>0.9$. This work will therefore provide benchmarks for further improvements of this detector.

At 250 GeV, ILC will provide far superior accuracies than LEP1. It will unambiguously allow to establish or exclude the presence of the LEP1 anomaly, without any ambiguity of sign given the $\gamma$-Z interference effect. Initial beam polarisation is mandatory in the process of separating the Z and photon components.

The presence of anomalous tensor couplings can significantly affect the angular distribution which would depart from the standard $1+\cos\theta^2$ behaviour. Using this feature, one can set model independent upper bounds on the tensor couplings $F_{2V}$ and $F_{2A}$.

b quark EW coupling measurements will be complementary to top quark coupling measurements given that both quarks pertain to the same weak isodoublet. This was illustrated by the EFT approach where the improvement in accuracy was nicely illustrated.

The reach in sensitivity on BSM physics surpasses the mass domain covered by LHC and therefore allows to anticipate on the relevance of future very high energy colliders.

# APPENDIX

In this section, one derives the formulae needed to extract the b quark EW couplings from the measured angular distribution. One follows the formalism of [9].

$$\frac{d\sigma}{d\cos\theta} = S_0\{(1+\cos^2\theta)(|F_{1V}+F_{2V}|^2+|F_{1A}'|^2) + \sin^2\theta(|\gamma^{-1}F_{1V}+\gamma F_{2V}|^2+|\gamma F_{2A}'|^2) - 4\cos\theta\, F_{1A}'(F_{1V}+\Re F_{2V})\}$$

where $F_{1A'}=\beta F_{1A}$, $F_{2A}'=\beta F_{2A}$ and $S_0=6\beta N_c\sigma_{pct}/8$ for a fully polarized beam, $N_c=3$ is the color factor, $\sigma_{pct}\sim 100\text{fb}/s_{TeV^2}$ being the so-called point-like cross section. For b quarks at 250 GeV, one has $S_0=3600$ fb and $\gamma\sim 125/5=25$.

In above expression, the $\sin^2\theta$ term is sensitive to the tensor terms $F_{2V}$ and $F_{2A}$, which are enhanced by the Lorentz factor $\gamma$. This feature was used in the 3-parameter fit to provide upper limits on these terms.

For the left handed polarisation, at the tree level, one has [9]:

$$F_{1V}^L = -Q_b + e_L BW \frac{-0.25 - Q_b s_W^2}{s_W c_W},\quad F_{1A}^L = e_L BW \frac{0.25}{s_W c_W},\quad BW = \frac{s}{s-M_Z^2},\quad e_L = \frac{-0.5+s^2 w}{cwsw},$$

$F_{1VL}=0.630$, $F_{1AL}=-.43$, $F_{1VR}=0.079$ and $F_{1AR}=0.367$.

This gives $\sigma_L/\sigma_{pct}=3.49$ and $\sigma_R/\sigma_{pct}=0.85$, again for fully polarized beams. One also derives $AFB_L=0.70$ and $AFB_R=0.28$. For effective polarisations, one has:

$$\sigma = 0.25(1-PP')(\sigma_R+\sigma_L) + 0.25(P-P')(\sigma_R-\sigma_L).$$

$F_{2A}$ is negligibly small within the SM. For $F_{2V}$, the QCD form factor is computed from [9]:

$F_{2Q}(\beta)=F_{2Q}(0)(2I_2)$, where $I_2 = \frac{1-\beta^2}{4\beta}\left[\ln\frac{1-\beta}{1+\beta}+i\pi\right]$, which gives $2I_2=-6\cdot 10^{-3}+i2.5\cdot 10^{-3}$.

$F_{2Q}(0)$ can be found, for the QCD part, in [15] up to the second order. The form factors for QCD (and for QED) do not depend on the EW coupling. Therefore, for these contributions, one can write $F_{2VL}(\beta)=F_{1VL}(\beta)F_{2Q}$. This of course would not be right for EW contribution and, eventually for BSM contributions. For this reason the results are given in terms of $F_{2VL}$ and $F_{2VR}$ instead of $F_{2VZ}$ and $F_{2Vem}$. Adding first and second order in QCD $F_{2Q}(0) =-3(15.3+4.7)10^{-3}=-0.06$ at the top mass scale. The second order QCD term is ~30% of the first order term. This gives:

$$\Re F_{2Q}=3.6\cdot 10^{-4},\ \Re F_{2VL}=2.3\cdot 10^{-4},\ \Re F_{2VR}=2.6\cdot 10^{-5}.$$

From these values one can easily deduce that for $e_{-L}$, the coefficient of $1-\cos^2\theta$ is $\sim 5.4\cdot 10^{-6}$ that is negligible compared to the coefficient of $1+\cos^2\theta$ and therefore we have ignored this term in our 2-parameter fit. The same conclusions apply for the right handed polarisation.

In practice one can neglect the quadratic terms $\Re F_{2V}^2$ and $\Im F_{2V}^2$ which leaves only the interference terms which, for QCD and QED, go like $\Re F_{2V}=F_{1V}\Re F_{2Q}$:

$$\frac{d\sigma}{d\cos\vartheta} = S_0\{(1+\cos^2\vartheta)(F_{1V}^2 + 2F_{1V}\Re F_{2V} + F_{1A}^2) - 4\cos\vartheta\, F_{1A}(F_{1V} + \Re F_{2V})\}$$

In the 2-parameter fit S(1+ cos$\theta_b^2$)+Acos$\theta_b$ , one has : S~ $F_{1V}^2 + 2F_{1V}\Re F_{2V} + F_{1A}^2$ and

A~ $-4F_{1A}(F_{1V} + \Re F_{2V})$ . Assuming that only $F_{2V}$ varies, one has: dA/A~$\Re F_{2VL}$/$F_{1VL}$=1.59$\Re F_{2VL}$, dS/S=2.17$\Re F_{2VL}$ for e-$_L$ and dA/A=13.5$\Re F_{2VR}$ ,dS/S=1.12$\Re F_{2VR}$ for e-$_R$ , which gives at the one standard deviation level:

$$\Re F_{2VL} < 0.0017 \,,\ \Re F_{2VR} < 0.003$$

If case of a 3-parameter fit S(1+ cos²$\theta_b$)+Acos$\theta_b$+Tsin²$\theta_b$ with T=$S_0$(|$\gamma^{-1}F_{1V}$+$\gamma F_{2V}$|²+|$\gamma F_{2A}'$|²) .

One finds, at the one standard deviation level, the following two bounds:

$$|F_{2VL}| < 0.005 \quad \text{and} \quad |F_{2VR}| < 0.011$$

also valid for $F_{2AL}$ and $F_{2AR}$.

**Acknowledgements:**

This work was supported within the 'Quarks and Leptons' programme of the CNRS/IN2P3, France. The results benefit from the enlightening discussions in the framework of the French-Japanese FJPPL/TYL 'virtual laboratory' on top physics, particularly through comments by Keisuke Fujii and François LeDiberder. We would like to thank Werner Bernreuther for providing his expertise on QCD corrections to static heavy quark form-factors. We acknowledge the precious help from the ILD physics and software working groups, in particular from F. Gaede and A. Miyamoto for efficiently providing high quality simulated data.